\documentclass[useAMS]{mn2e}
\usepackage{times,natbib}
\usepackage{multirow}
\usepackage[dvips]{graphicx}
\font \bolditalics = cmmib10
\def\bx#1{\leavevmode\thinspace\hbox{\vrule\vtop{\vbox{\hrule\kern1pt
        \hbox{\vphantom{\tt/}\thinspace{\bf#1}\thinspace}}
      \kern1pt\hrule}\vrule}\thinspace}
\def \vc #1{{\textfont1=\bolditalics \hbox{$\bf#1$}}}

\def\phig{{\vc \phi}}

\def\thetag{{\vc \theta}}

\def\gammag{{\vc \gamma}}

\def\psig{{\vc \psi}}

\def\be{\begin{equation}}
\def\ee{\end{equation}}
\def\ba{\begin{eqnarray}}
\def\ea{\end{eqnarray}}

\def\ltsima{$\; \buildrel < \over \sim \;$}
\def\lsim{\lower.5ex\hbox{\ltsima}}
\def\gtsima{$\; \buildrel > \over \sim \;$}
\def\gsim{\lower.5ex\hbox{\gtsima}}

\begin{document}
\title[]{Shear and Magnification: Cosmic Complementarity}

\author[Van Waerbeke]
  {\parbox[]{6.in}{L. Van Waerbeke\\
  \footnotesize University of British Columbia, 6224 Agricultural Road,
    Vancouver, V6T 1Z1\, B.C.,Canada.\\
}}

\date{11 January 2007}

\maketitle

\begin{abstract}
The potential of cosmic shear to probe cosmology is well recognized
and future optical wide field surveys are currently being designed
to optimize the return of cosmic shear science. High precision
cosmic shear analysis requires high precision photometric redshift.
With accurate photometric redshifts, it becomes possible to measure
the cosmic magnification on galaxies {\it by} galaxies and use it as
a probe of cosmology. This type of weak lensing measurement will not
use galaxy shapes, instead it will strongly rely on precise
photometry and detailed color information. In this work it is shown
that such a measurement would lead to competitive constraints of the
cosmological parameters, with a remarkable complementarity with
cosmic shear. Future cosmic shear surveys could gain tremendously
from magnification measurements as an independent probe of the dark
matter distribution leading to a better control of observational and
theoretical systematics when combined with shear.
\end{abstract}


\section{Introduction}

The distortion of the images of distant galaxies due to mass
inhomogeneities along the line-of-sight, the cosmic shear, is a
powerful tool to probe the dark matter distribution in the Universe.
Recent results (\cite{benjamin2007},\cite{fu2008}) demonstrate that
the technique of cosmic shear has greatly matured since the early
detections. For recent reviews, see \cite{munshietal2008} and
\cite{hoekstrajain2008}. Nevertheless, cosmic shear still suffers
from practical difficulties encountered with galaxy shape
measurements, those difficulties triggered a massive effort aiming
at a better control of the residual systematics (\cite{heymans2006},
\cite{massey2007}, \cite{great08}). It becomes increasingly
important to explore parallel avenues to cosmic shear which could
provide a similar, yet independent, probe of dark matter
distribution and give additional ways for controlling the cosmic
shear residual systematics. The detection of cosmic magnification
from the Sloan Digital Sky Survey (\cite{scr2005}) could provide an
interesting and viable option. The gravitational lensing effect
leads to a change in the object size through the {\it
magnification}, and since the surface brightness is conserved, the
apparent magnitude of distant galaxies is modified. This effect
leads to a non vanishing angular cross-correlation between distant
(lensed) objects and foreground (lenses) galaxies, which can serve
as a probe of cosmology (\cite{n1989},\cite{mj98}).

The cosmic magnification has been measured successfully in the Sloan
Digital Sky Survey through the QSO-galaxy correlation function
(\cite{scr2005}). Previous attempts to measure magnification faced a
challenging cosmological interpretation, but \cite{scr2005} showed
for the first time convincing evidence of the cosmological origin of
the signal. The main challenge in the measurement of the
magnification effect lies in the fact that the foreground and
background populations have to be clearly separated in redshift in
order to avoid any spurious angular clustering contamination. This
was possible with the SLOAN survey because bright QSOs at redshift
higher than $1.5$ were clearly distinguishable from low redshift
galaxies. This was essentially the main reason for the successful
detection in \cite{scr2005}. Ideally, we would like to use galaxies
as sources because they are much more numerous and they spread over
a larger redshift range than QSOs. It would probe similar dynamical
regimes and redshift range as cosmic shear.

The tremendous progress made in photometry and photometric redshift
measurements in the last decade
(\cite{oetal2006},\cite{whb2008},\cite{erbenetal2009},\cite{couponetal2009})
may enable this possibility. In particular, future wide field
optical surveys \footnote{e.g. Large Synoptic Survey Telescope
http://www.lsst.org/lsst and Joint Dark Energy Mission
http://jdem.lbl.gov/DarkEnergy.html for ground based and space based
surveys respectively} will be designed to cover a wide range of
wavelengths with many filters in order to obtain accurate
photometric redshifts for lensing and many other applications. The
required accuracy (\cite{huterer2006}) is such that it enables the
possibility to measure the magnification effect on galaxies by
galaxies. This would open an entirely new window for dark matter
searches which could become a very useful complement of cosmic shear
measurements. Cosmic magnification has several advantages over
cosmic shear which will be discussed in this paper, but a
particularly noticeable one is that magnification is not sensitive
to Point Spread Function corrections. The magnification offers a
complementary approach to cosmic shear, with a radically different
sensitivity to observational and theoretical systematics. This paper
is an exploratory work which examines to what extends the cosmic
magnification could be measured on galaxies and how and why it could
be integrated in the design of future ambitious surveys.

The first section introduces notations and definitions used in this
work. Section 3 details the calculation of the cosmic magnification
covariance matrix necessary in order to gauge the ability of cosmic
magnification to measure cosmological parameters and Section 4
details how the CFHTLS Deep data are used to calibrate the free
parameters needed in the predictions. In Section 5 we discuss the
complementarity with cosmic shear and how magnification can be used
to constrain cosmology. We then conclude on the outcome and
implications of this work.

\section{Theory}
\subsection{Definitions and Notations}

The gravitational lensing effect is to first order described by the
magnification matrix $A$

\begin{equation}
A=\pmatrix{1-\kappa+\gamma_1&-\gamma_2\cr
-\gamma_2&1-\kappa-\gamma_1}
\end{equation}
where $\kappa$ is the convergence and $\gammag=\gamma_1+i\gamma_2$
is shear in complex notation. In the following the notations and
definitions from \cite{svwjk1998} are being used. The convergence is
the line-of-sight integral

\begin{equation}
\kappa(\thetag)={3\over 2}\Omega_m\int_0^{w_H}{\rm d}~w g(w)
f_K(w){\delta(f_K(w)\thetag,w)\over a(w)},\label{kappadef}
\end{equation}
where $w(z)$ is the comoving radial coordinate at redshift $z$,
$f_K(w)$ is the angular diameter distance at distance $w$, and
$\delta(f_K(w)\thetag,w)$ is the mass density contrast. $w_H$ is the
horizon distance and $g(w)$ is an integral over the source redshift
distribution:

\begin{equation}
g(w)=\int_w^{w_H}{\rm d} w' p_w(w'){f_K(w'-w)\over f_K(w')},
\end{equation}
where $p_w(w){\rm d}w=p_z(z){\rm d}z$ is the source redshift
distribution.

Let us consider a lensing survey complete down to some limiting
magnitude $m_{\rm lim}$. Along a given line-of-sight $\thetag$, the
number density of unlensed galaxies with magnitude $[m,m+dm$ is
defined by $N_0(m)dm$ and $N(f,\thetag)dm$ is the number of lensed
galaxies. Calling $\mu$ the magnification factor, then the galaxy
number counts follow the relation \cite{n1989}:

\begin{equation}
N(m,\thetag)dm=\mu^{2.5 s(m)-1}N_0(m)dm, \label{lensedcounts}
\end{equation}
where $s(m)$ is the slope of the galaxy number counts at magnitude
$m$:

\begin{equation}
s(m)={{\rm d}N_0(m)\over {\rm d}m}
\end{equation}
To first order, when the shear and convergence fields are small
compared to one $(\kappa$,$\gamma$ $<<1$), the magnification is
given by

\begin{equation}
\mu(\thetag)\simeq 1+2\kappa(\thetag). \label{mudef}
\end{equation}
The magnification can be measured through the angular
cross-correlation of a background lensed population of galaxies with
foreground galaxies (the lenses). The angular cross-correlation
$w_{12}(\theta)$ at separation $\theta=|\thetag_i-\thetag_j|$ is
given by

\begin{equation}
w_{12}(\theta)=\langle
\delta_1(\thetag_i)\delta_2(\thetag_j)\rangle,
\end{equation}
where $\delta_1$ and $\delta_2$ are the fractional densities of the
foreground and background galaxy populations at redshift $z_1$ and
$z_2$. Expressing the fractional densities as fractional number
counts and taking into account the effect of lensing by large scale
structures (Eq.\ref{lensedcounts}), the cross-correlation at angular
separation $\theta$ becomes (\cite{b1995}):

\begin{eqnarray}
w_{12}(\theta)&=&3 b (\alpha(m)-1)\Omega_m \int {s{\rm d}s\over
2\pi}\int{\rm d}w {g(w)\over a(w)}{p_1(w)\over f_K(w)}\times\cr &&
P_{3D}\left({s\over f_K(w)};w\right) J_0(\theta s). \label{2ptstat}
\end{eqnarray}
where $P_{3D}$ is the 3-dimensional matter power spectrum,
$J_0(\theta s)$ the zeroth order Bessel function of the first kind,
and $\alpha(m)=2.5s(m)$. The galaxy basing parameter $b$ is defined
by $\delta_1=b\delta$. Eq.\ref{2ptstat} shows the well known fact
that the magnification effect depends on the number count slope of
the lensed population and on how the foreground galaxies trace the
underlying matter distribution. In the following we will take $b=1$,
although means to alleviate this assumption or to measure the
biasing will be discussed in the conclusion. This assumption does
not alter the main conclusion of this work regarding the
cosmological constraints of the cosmic magnification.

\subsection{Practical Estimator}

We assume we have a compact survey of area $A$, with two galaxy
populations at two different redshifts $z_1$ and $z_2$ and number
density of galaxies $n_1$ and $n_2$ respectively. The sky is finely
gridded with "pixels" at location $\thetag_i$, and the "density" of
galaxy population $k$ at this location $\thetag_i$ is called
$\delta_k(\thetag_i)$. Note that for a sufficiently high resolution
grid, $\delta_k$ is zero or one.

The angular cross-correlation between the two galaxy populations,
for angular separation $\theta$, is given by

\begin{equation}
w_{12}(\theta)=\langle
\delta_1(\thetag_i)\delta_2(\thetag_j)\rangle_{\left|\thetag_i-\thetag_j\right|=\theta}
\label{wdef}
\end{equation}
The discretized estimator of this correlation function is given by

\begin{equation}
w_{12}(\theta)={1\over
N_p(\theta)}\sum_{i,j}\delta_1(\thetag_i)\delta_2(\thetag_j)\Delta_\theta(ij)
\label{westimator}
\end{equation}
where $N_p(\theta)$ is the number of pairs with separation
$[\theta-d\theta/2,\theta+d\theta/2]$ and $\Delta_\theta(ij)$ is
defined as
$\Delta_\theta(ij)=\Delta_\theta(\left|\thetag_i-\thetag_j\right|)=1$
for

$\theta-d\theta/2<\left|\thetag_i-\thetag_j\right|<\theta+d\theta/2$
and zero otherwise. Since we assumed a continuous survey, the number
of pairs $N_p(\theta)$ is given by:

\begin{equation}
N_p(\theta)\simeq A~n_1~2\pi\theta~d\theta~n_2\label{Npdef}
\end{equation}
As a self consistency check, we should demonstrate that
Eq.\ref{westimator} is an unbiased estimator of Eq. \ref{wdef}. With
a number of galaxies $N_1=n_1A$ and $N_2=n_2A$, there are
$N_1(N_2-1)\sim N_1N_2$ pairs, so for large $N$, and applying the
change of variable $\psig=\thetag_1-\thetag_2$, Eq.\ref{westimator}
can be rewritten:

\begin{eqnarray}
w_{12}(\theta)&=&{1\over N_p(\theta)}{N_1N_2\over A^2}\int_A{\rm
d}\thetag_1\int_A{\rm d} \thetag_2
\Delta_\theta(12)\langle\delta_1(\thetag_1)\delta_2(\thetag_2)\rangle\cr
&=&{1\over N_p(\theta)}{N_1N_2\over A^2}\int{\rm d}\thetag_1\int{\rm
d} \thetag_2 \Delta_\theta(12)
w_{12}(\left|\thetag_1-\thetag_2\right|)\cr &=&{1\over
N_p(\theta)}{N_1N_2\over A^2}\int_A{\rm d}\thetag_1\int_0^{2\pi}{\rm
d}\hat\psi \int_{\theta-d\theta/2}^{\theta+d\theta/2}\psi{\rm d}
\psi ~w_{12}(\psi)\cr &\simeq&{1\over N_p(\theta)}{N_1N_2\over
A}2\pi \int_{\theta-d\theta/2}^{\theta+d\theta/2}\psi{\rm d} \psi
~w_{12}(\theta)\cr &\simeq& w_{12}(\theta)
\end{eqnarray}
This shows that Eq.\ref{westimator} is an unbiased estimator of
Eq.\ref{wdef}. A practical implementation of the estimator
Eq.\ref{westimator} is given in \cite{ls1993} for which
Eq\ref{westimator} can be rewritten as:

\begin{equation}
w_{12}={(N_1-R_1)(N_2-R_2)\over R_1 R_2}
\end{equation}
where $R_1$ and $R_2$ are the number of {\it objects} in two
distinct random samples. This estimator follows directly from the
explicit expressions for $\delta_1$ and $\delta_2$:

\begin{equation}
\delta_k={N_k-R_k\over R_k}
\end{equation}
where $k=(1,2)$ refers to the population $1$ or $2$.

\section{Covariance matrix}

The cosmic magnification covariance matrix is derived in this
section. It is also compared to the cosmic shear covariance matrix
and the efficiency of the two weak lensing techniques is compared.

The cosmic magnification signal is the cross-correlation function
$w_{12}(\theta)$ as defined in Eq.\ref{2ptstat}. The covariance
matrix is defined as $C_{\theta \phi}=\langle
w_{12}(\theta)w_{12}(\phi)\rangle$. With the practical estimator for
$w_{12}$ given by Eq.\ref{westimator}, the covariance matrix can be
written as:

\begin{eqnarray}
C_{\theta\phi}&=&{1\over
N_p(\theta)~N_p(\phi)}\sum_{ijkl}\Delta_\theta(ij)\Delta_\phi(kl)\times
\cr &&\langle\delta_1(\thetag_i)\delta_2(\thetag_j)
\delta_1(\thetag_k)\delta_2(\thetag_l)\rangle \label{Cdef}
\end{eqnarray}
$\delta_1(\thetag)$ and $\delta_2(\thetag)$ refer to the number
count contrast of the foreground and background populations
respectively. From the expressions for the magnification effect (Eqs
\ref{lensedcounts} and \ref{mudef}), the count density contrasts are
given by:

\begin{eqnarray}
\delta_2(\thetag)&=&{N_2(m,\thetag)-\overline{N_2(m)}\over
\overline{N_2(m)}}=2(\alpha(m)-1)\kappa(\thetag)+\delta_{N_2}(\thetag)\cr
\delta_1(\thetag)&=&{N_1(m,\thetag)-\overline{N_1(m)}\over
\overline{N_1(m)}}=\delta_{N_1}(\thetag) \label{deltadef}
\end{eqnarray}
where $\delta_{N_1}$ and $\delta_{N_2}$ are the unlensed count
density contrast. It is assumed that only the background population
is lensed, although in principle the foreground population contrast
$\delta_1(\thetag)$ should also contain a convergence term coming
from lenses at even lower redshift. This term is however negligible
for a reason given later in this Section (Eq. \ref{piexpr}). The
covariance matrix $C_{\theta\phi}$ contains therefore terms with
convergence and terms with unlensed density contrasts
$\delta_{N_i}$. The later is a major source of noise for the
magnification measurement since it depends on the clustering of
galaxies in each population (low and high redshift). The covariance
matrix can therefore be split in two terms:

\begin{equation}
C_{\theta\phi}=C_{\theta\phi}^S+C_{\theta\phi}^L+C_{\theta\phi}^P,
\end{equation}
Where $C^S$ is the clustering term, $C^L$ contains the weak lensing
contribution and $C^P$ is the shot noise Poisson term. We focuss on
$C^S$ first, before showing that the terms containing lensing
quantities $C^L$ is negligible in comparison. $C^P$ is calculated at
the end of this Section.

It is assumed that the galaxies distributions are gaussian random
fields, therefore four point correlation functions can be expressed
as the sum of product of all possible combinations of two-point
correlation functions. Moreover, the cross-correlation terms
$\langle \delta_{N_1}\delta_{N_2}\rangle$ are zero if the background
and foreground populations are well separated in redshift. The only
contribution to the clustering term in the covariance matrix will
come from the auto-correlation functions of each population, which
are defined as:

\begin{equation}
w_{kk}(\theta)=\langle
\delta_{N_k}(\thetag_i)\delta_{N_k}(\thetag_j)\rangle
\end{equation}
where $k=(1,2)$ refers to the population redshift. The clustering
term in the covariance matrix can therefore we written as

\begin{eqnarray}
C_{\theta\phi}^S&=&{1\over
N_p(\theta)~N_p(\phi)}\sum_{ijkl}\Delta_\theta(ij)\Delta_\phi(kl)
~\langle\delta_{N_1}(\thetag_i)\delta_{N_1}(\thetag_k)\rangle~\times\cr
&&\langle \delta_{N_2}(\thetag_j)\delta_{N_2}(\thetag_l)\rangle
\label{Cclustering}
\end{eqnarray}
The rest of the calculations follow closely the steps taken for the
shear covariance matrix in \cite{svwkm2002}. There is approximately
$\sim N_1^2N_2^2$ identical terms in Eq.\ref{Cclustering}, therefore
we can take the continuum limit. For clarity let's introduce the
quantity $C'=C\times N_p(\theta)~N_p(\phi)$, which can then be
written as:

\begin{eqnarray}
C'^S_{\theta\phi}&=& {N_1^2N_2^2\over A^4} \int_A{\rm d}\thetag_1
\int_A{\rm d}\thetag_2 \int_A{\rm d}\thetag_3 \int_A{\rm
d}\thetag_4~\Delta_\theta(12)\Delta_\phi(34)~\times\cr &&\cr
&&w_{11}(\left|\thetag_1-\thetag_3\right|)
~w_{22}(\left|\thetag_2-\thetag_4\right|)\cr &&\cr
 &=& {N_1^2N_2^2\over
A^3} \int_A{\rm d}\psig
\int_{\theta-d\theta/2}^{\theta+d\theta/2}\phi_1{\rm d}\phi_1
\int_0^{2\pi}{\rm d}\hat\phi_1
\int_{\phi-d\phi/2}^{\phi+d\phi/2}\phi_2{\rm d}\phi_2\times\cr
&&\int_0^{2\pi}{\rm d}\hat\phi_2~w_{11}(\left|\psig-\phig_1\right|)
~w_{22}(\left|\psig+\phig_2\right|)
\end{eqnarray}
where one of the four integrals over the survey area was carried
out, and the following change of variables has been applied:

\begin{eqnarray}
\thetag_2&=&\thetag_1+\phig_1\cr \thetag_4&=&\thetag_3+\phig_2\cr
\psig&=&\thetag_3-\thetag_1
\end{eqnarray}
We now define the following vectors:

\begin{eqnarray}
\psig_a&=&\left(\matrix{\psi\cos(\hat\psi)-\theta\cos(\hat\phi_1)\cr
\psi\sin(\hat\psi)-\theta\sin(\hat\phi_1)}\right)\cr \cr\cr
\psig_b&=&\left(\matrix{\psi\cos(\hat\psi)-\phi\cos(\hat\phi_2)\cr
\psi\sin(\hat\psi)-\phi\sin(\hat\phi_2)}\right)\nonumber
\end{eqnarray}
Then the final step is relatively straightforward to carry out, and
for the covariance matrix we get:

\begin{equation}
C^S_{\theta\phi}={2\over\pi~A}\int_0^\infty \psi{\rm
d}\psi~\int_0^\pi{\rm
d}\hat\phi_1~w_{11}(\left|\psig_a\right|)\int_0^\pi{\rm
d}\hat\phi_2~w_{22}(\left|\psig_b\right|) \label{Cmatrix}
\end{equation}
This term describes the contribution to the comic magnification
covariance matrix from the intrinsic clustering of the background
and foreground populations. In the following section, estimates and
comparison with the shear covariance matrix will be given. Let us
now consider the lensing-clustering mixed terms in Eq. \ref{Cdef}.
From the definition of the number count density contrast in Eq.
\ref{deltadef}, the mixed term can be written as:

\begin{equation}
C^L_{\theta\phi}={1\over
N_p(\theta)~N_p(\phi)}\sum_{ijkl}\Delta_\theta(ij)\Delta_\phi(kl)
~\Pi_{ijkl}
\end{equation}
with

\begin{eqnarray}
\Pi_{ijkl}&=&
\langle\delta_{N_1}(\thetag_i)\delta_{N_1}(\thetag_k)\rangle \langle
\kappa(\thetag_j)\kappa(\thetag_l)\rangle\cr
&+&\langle\delta_{N_1}(\thetag_i)\kappa(\thetag_l)\rangle \langle
\kappa(\thetag_j)\delta_{N_1}(\thetag_k)\rangle\cr
&+&\langle\delta_{N_1}(\thetag_i)\kappa(\thetag_j)\rangle \langle
\kappa(\thetag_l)\delta_{N_1}(\thetag_k)\rangle \label{piexpr}
\end{eqnarray}
Following the same calculations as for the clustering term, the
first term in Eq.\ref{piexpr} leads to the following contribution to
the covariance matrix:

\begin{equation}
C^L_{\theta\phi}={2\over\pi~A}\int_0^\infty \psi{\rm
d}\psi~\int_0^\pi{\rm
d}\hat\phi_1~w_{11}(\left|\psig_a\right|)\int_0^\pi{\rm
d}\hat\phi_2~\xi_\kappa(\left|\psig_b\right|) \label{Cmix}
\end{equation}
where $\xi_\kappa$ is the convergence correlation function (which is
equal to the shear correlation function in the weak lensing limit).
This expression is very similar to the clustering term in
Eq.\ref{Cmatrix}, except that the auto-correlation function $w_{22}$
has been replaced by $\xi_\kappa$. For that reason, Eq.\ref{Cmix}
leads to a negligible contribution to the cosmic magnification
covariance matrix, because $\xi_\kappa$ is always orders of
magnitude smaller than $w_{22}$. For the same reason, the last two
terms in Eq.\ref{piexpr} can be neglected, as well as the
contribution from the lensing of the foreground population, which
was previously neglected in Eq.\ref{deltadef}. This leads to the
surprising result that the sampling variance from the lensing effect
for the cosmic magnification is negligible compared to the source of
statistical noise (the intrinsic clustering of the foreground and
background populations). This is opposite to the cosmic shear
situation, where the sampling variance dominates the source of
statistical noise (the intrinsic ellipticity) for scales above a few
arcminutes (\cite{svwkm2002}), even when non-gaussian errors are
considered (\cite{semetal2007}).

The derivation of Eq.\ref{Cmatrix} assumed that the Poisson error
resulting from discrete sampling could be neglected. This is no
longer the case if the number density of galaxies becomes too small,
it is therefore necessary to add the contribution from sampling
noise to the covariance matrix $C^P$. This can be done by describing
the sampling by a Poisson point process, which leads to a Poisson
error for each angular bin of the cross-correlation function
$w_{12}(\theta)$. For a bin of width ${\rm d}\theta$ at angular
separation $\theta$, the error $\Delta w_{12}(\theta)$ on the
cross-correlation is given by the inverse square root of the number
of pairs $N_p(\theta)$ contributing to that bin (see Eq.
\ref{Npdef}).

\section{CFHTLS Deep survey}

Our ability to measure cosmic magnification depends on the amount of
clustering in the foreground and background galaxy samples, and on
the number count slope $\alpha(m)$. The CFHTLS Deep survey provides
us with the necessary ingredients to calculate the magnification
covariance matrix. The survey is complete down to $i'\simeq 25.5$
and with four independent fields of one square degree each, the
impact of sampling variance on the angular clustering will be
limited. It therefore provides a perfect sample which could be
applied to shallower surveys such as the CFHTLS Wide. The Deep
survey has also the unique advantage to provide accurate photometric
redshift estimates (from \cite{oetal2006}), which can be used for
the foreground and background separation.

\begin{figure}
\begin{center}
\resizebox{3.0in}{!}{\includegraphics{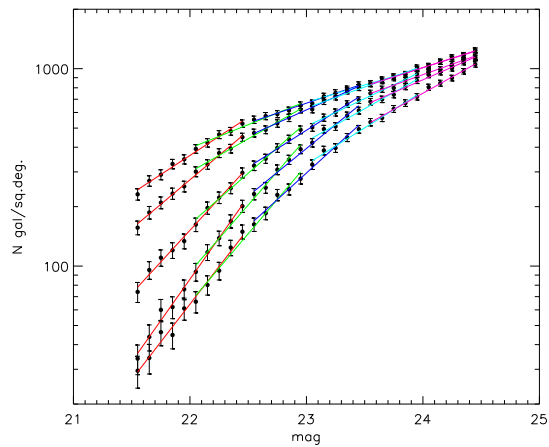}}
\end{center}
\caption{Number counts $N(m)$ for difference redshift slices.
Poisson errors are indicated and the straight line shown for ech
magnitude bin have a slope $\alpha(m)-1$. From top to bottom the
redshift slices are $z=[1.1,1.4]; [1.0,1.3]; [0.9,1.2]; [0.8,1.1];
[0.7,1.0]$. The top, middle and bottom lines correspond to the
magnitude bin and redshift slices listed in Table \ref{tab:counts}.}
\label{fig:counts}
\end{figure}

The CFHTLS Deep data have been divided in $5$ redshift bins and $5$
magnitude bins within the magnitude range $[21.5,24.5]$. A faint
magnitude cut of $i'=24.5$ corresponds to the limiting magnitude of
the CFHTLS Wide, therefore the results presented here can be applied
to this survey too. The slope of the number counts has been measured
for each redshift slice for $5$ different magnitude bins. The result
is shown in Table \ref{tab:counts} for the lowest, intermediate and
highest redshift bins, and Figure \ref{fig:counts} shows the fitted
slope for the background populations for all bins.

\begin{table}
\centering
\begin{tabular}{l l l l}
\hline
Redshift slice & $[0.7,1.0]$&$[0.9,1.2]$ & $[1.1,1.4]$\\
\hline
$m=[21.5,22.5]$&  -0.03 & 0.60 & 0.91 \\
 & 1.04 & 0.46 & 0.20 \\
 \hline
$m=[22,23]$&  -0.41 & 0.29 & 0.74 \\
 &  1.46 & 0.87 & 0.45 \\
 \hline
$m=[22.5,23.5]$&  -0.50  & -0.07 & 0.35 \\
 &  1.87  & 1.40 & 0.87 \\
\hline
$m=[23,24]$& -0.56 & -0.30 & -0.08 \\
 &  2.31 & 2.00 & 1.43 \\
  \hline
$m=[23.5,24.5]$&  -0.54  & -0.47 & -0.18 \\
 &  2.83 & 2.61 & 2.13 \\
\end{tabular}
\caption{Table showing for three redshift slices and each magnitude
bin the measured value of $\alpha(m)-1$ and the number density
$n_{\rm gal}$ of galaxies per arcmin$^2$. For each magnitude bin,
the top row shows $\alpha(m)-1$ and the bottom row $n_{\rm gal}$.}
\label{tab:counts}
\end{table}

\begin{figure}
\begin{center}
\resizebox{3.0in}{!}{\includegraphics{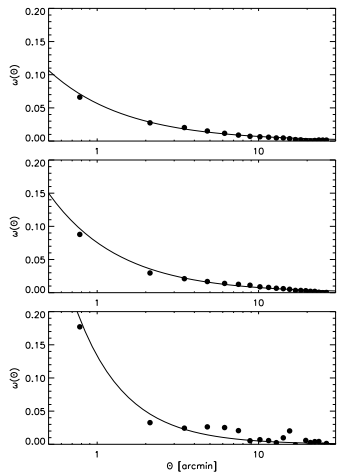}}
\end{center}
\caption{Auto-correlation functions measurement and power law fits.
Top panel: $z=[1.1,1.4]$ and $m=[21.5,22.5]$; middle panel:
$z=[0.7,1.0]$ and $m=[23.5,24.5]$; bottom panel: foreground galaxy
population with $z=[0.1,0.6$ and $m=[21.5,24.5]$.}
\label{fig:powerfit}
\end{figure}

The autocorrelation functions are necessary in order to estimate the
covariance matrix of the cosmic magnification. The standard power
law model has been used to fit the correlation functions for the
different galaxy populations. Figure \ref{fig:powerfit} shows the
fit for the three background populations listed in Table
\ref{tab:counts}. Although the amplitude and shape agree with
previous measurements in the literature (\cite{hmc2008}) the
integral constraint has not been determined here. This should only
lead to a minor change in the covariance matrix since most of the
contribution come from the small angular scales (less than 10
arcminutes).

\section{\label{compare}Complementarity with cosmic shear}

\subsection{Signal-to-noise}

The cosmic magnification and shear have different sources of
sampling and statistical noise and this section is devoted to the
signal-to-noise comparison of the two statistics for the same galaxy
population. We compare magnification to the cosmic shear top-hat
variance, defined as:

\begin{equation}
\langle\gamma^2\rangle={9\over 2\pi}\Omega_m^2 \int s{\rm d}s
\int{\rm d}w {g^2(w)\over a^2(w)} P_{3D}\left({s\over
f_K(w)};w\right) {J_1^2(\theta s)\over (s\theta)^2}
\label{shear2pts}
\end{equation}
where $J_1(\theta s)$ is the first order Bessel function of the
first kind. The corresponding covariance matrix has been calculated
using the equation derived in \cite{svwkm2002}. We consider two
different background galaxy populations, at $z=[0.7,1.0]$ and
$z=[1.1,1.4]$. The number count used for the cosmic shear is the
total number count for that redshift slice summed over all magnitude
bins. For the magnification, the chosen magnitude range is
$m=[23.5,24.5]$ and $m=[21.5,22.5]$ for the low and high redshift
background respectively. The foreground population is at
$z=[0.1,0.6]$ with $m=[21.5,24.5]$. The corresponding $\alpha(m)$
and number count densities are shown in Table \ref{tab:counts}.

\begin{figure}
\begin{center}
\resizebox{3.3in}{!}{\includegraphics{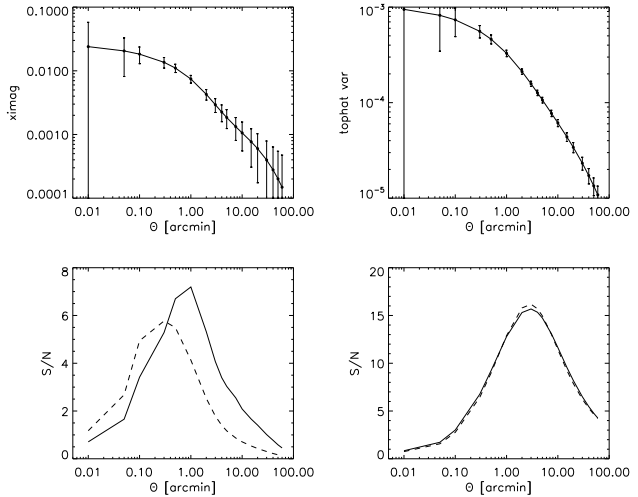}}
\end{center}
\caption{Top panels show the signal of the cosmic shear (right plot)
and cosmic magnification (left plot). The error bars include the
statistical noise and sampling variance for both techniques for a
survey of $400$ square degrees. For the cosmic magnification (left
panel) the foreground population is chosen with $z=[0.1,0.6]$ and
$m=[21.5,24.5]$. The source redshifts are $z=[1.1,1.4]$ (solid line)
and $z=[0.7,1.0]$ (dashed line). Galaxy number densities as
indicated in Table \ref{tab:counts}.} \label{fig:THplot}
\end{figure}

\begin{figure}
\begin{center}
\resizebox{3.3in}{!}{\includegraphics{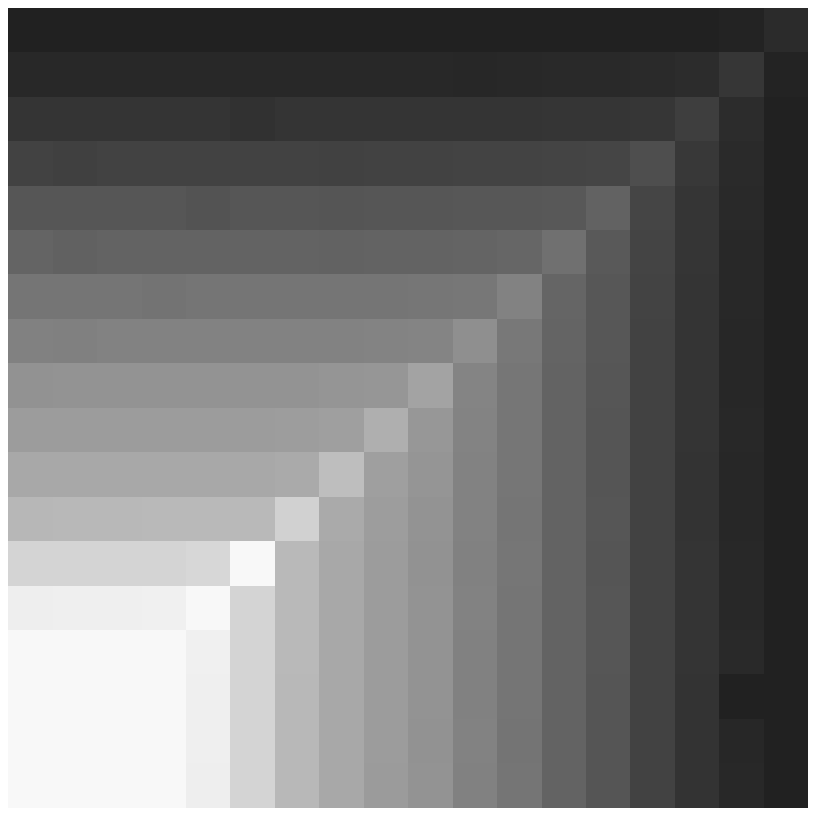}\includegraphics{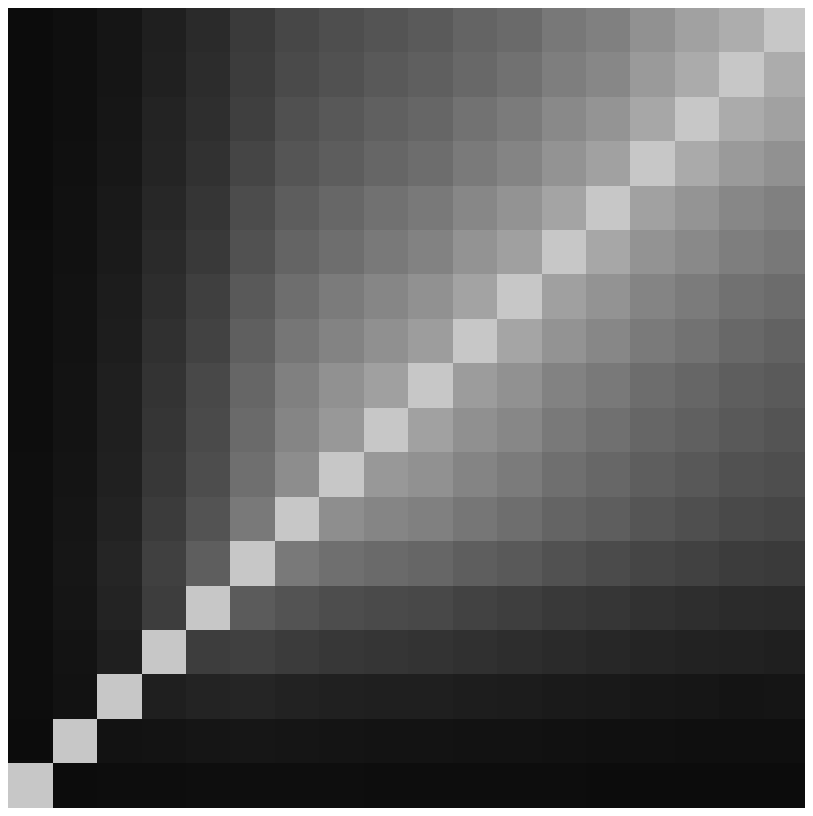}}
\end{center}
\caption{The left panel shows the covariance matrix of the cosmic
magnification for the population of galaxies used in Figure
\ref{fig:THplot}. The right panel shows the corresponding
correlation matrix. The smallest scale is $0.01$ arcminutes (bottom
left) and the largest scale is $90$ arcminutes (top right).}
\label{fig:Cmatrix}
\end{figure}

Figure \ref{fig:THplot} shows the signal-to-noise as function of
scale for the cosmic magnification and shear. It is very interesting
that using similar galaxy populations, shear and magnification are
measured with similar precision. There is only a factor 2
difference, which can be further reduced from the use of combined
magnitude slices for the magnification for the same source redshift,
although it is not the goal of this work to investigate optimal
estimators of cosmic magnification. The signal-to-noise difference
between magnification and shear shown in Figure \ref{fig:THplot} is
in agreement with the results from ray-tracing simulations
(\cite{th2003}). Figure \ref{fig:Cmatrix} shows that the intrinsic
clustering in the cosmic magnification covariance matrix leads to
significant cross-correlation between scales which dominates at
large angular separation. This is why the cosmic magnification
signal-to-noise decreases for scales larger than $10$ arc-minutes,
where Poisson noise is negligible (see Figure \ref{fig:THplot}).


\subsection{Cosmological constraints}

\begin{figure}
\begin{center}
\resizebox{3.0in}{!}{\includegraphics{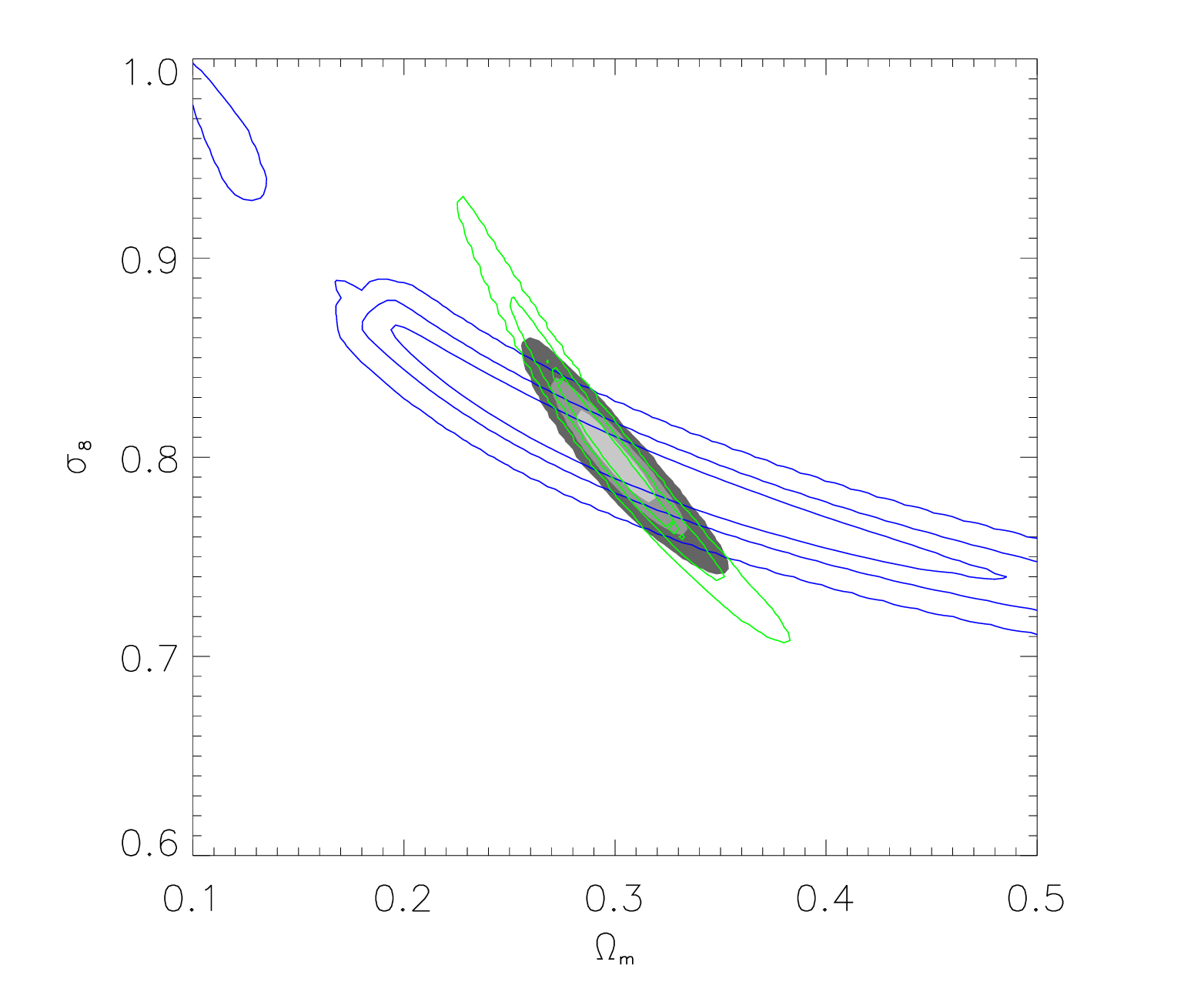}}
\resizebox{3.0in}{!}{\includegraphics{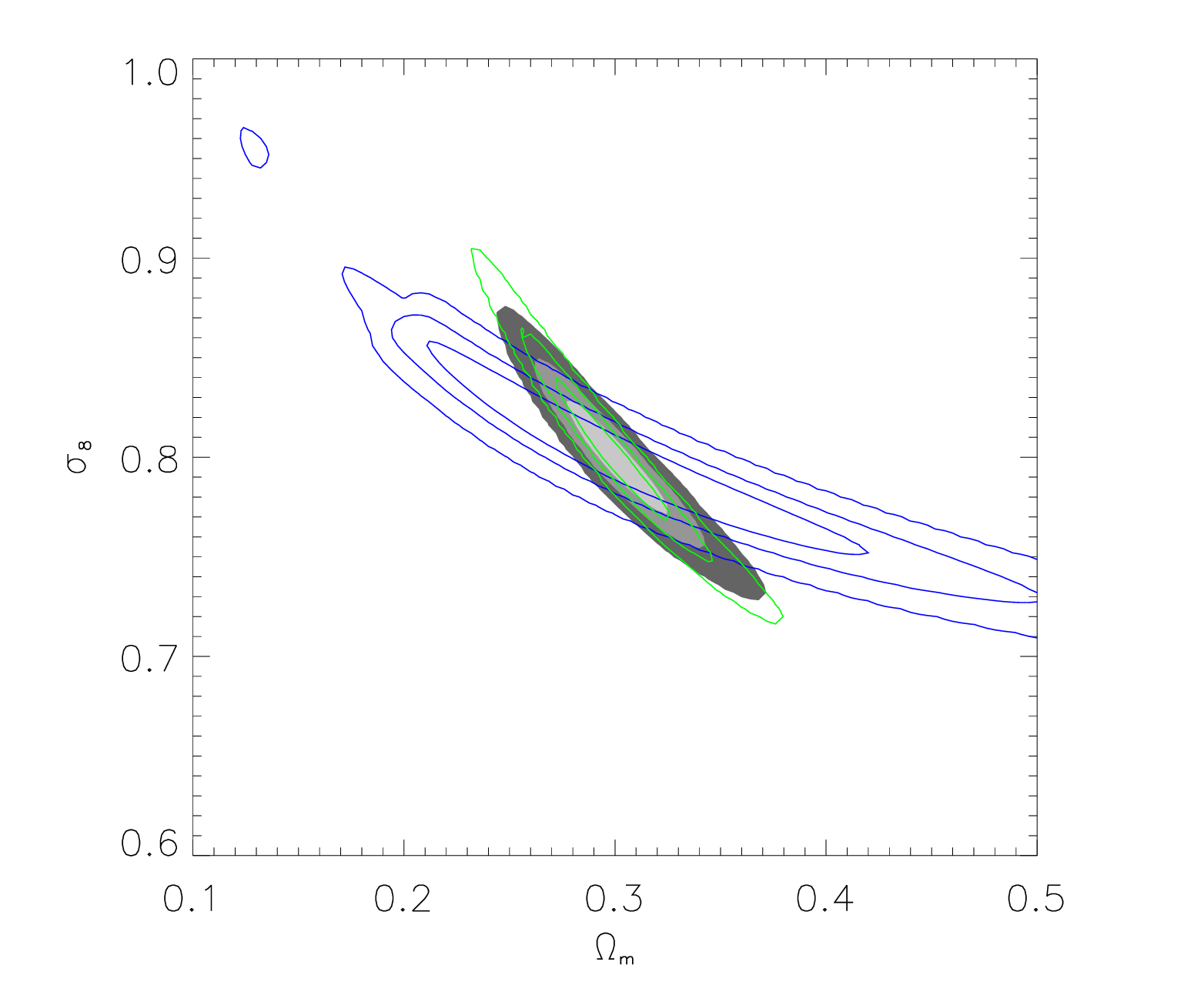}}
\end{center}
\caption{Cosmological parameters constraints from shear and
magnification for a 1500 sq.deg. survey. Top panel: the
magnification is measured on the $z=[0.7,1.0]$ and $m=[23.5,24.5]$
with the foreground galaxies located at $z=[0.1,0.6]$; the shear is
measured on the $z=[0.7,1.0]$ and $m=[21.5,24.5]$ galaxies. Bottom
panel: the magnification is measured on the $z=[1.1,1.4]$ and
$m=[21.5,22.5]$; the shear is measured on the $z=[1.1,1.4]$ and
$m=[21.5,24.5]$ galaxies. The filled contour shows the error contour
obtains from a $750$ sq. deg. shear analysis combined with a $750$
sq.deg. magnification analysis.} \label{fig:contours}
\end{figure}

In order to test the ability of cosmic magnification to constrain
the cosmological parameters, we take $b=1$ and the likelihood of
cosmological models for a range of values of $\Omega_m$ and
$\sigma_8$ are calculated. A fiducial model is chosen with
$\Omega_m=0.3$, $\Omega_\Lambda=0.7$ and $\sigma_8=0.8$, and the
covariance matrix given by Eq. \ref{Cmatrix} including the shot
noise is calculated. The foreground redshift slice is chosen as
$z=[0.1,0.6]$ and the background population is the same as for the
two cases shown in Figure \ref{fig:THplot}. The survey area is
$A=1500$ square degrees. For the comparison, the shear has been
calculated using the same background population as for the
magnification but with no magnitude restriction (therefore shot
noise is decreased compared to magnification). The results are shown
in Figure \ref{fig:contours}. As it was anticipated in the previous
section, the performance of the two techniques is relatively
similar. Note that a 50\% of the area (750 sq.deg.) used for
magnification and 50\% used for shear performs better than 100\% of
the area used for any of the shear or magnification taken alone. One
should also note that degeneracy-breaking in the
$\Omega_m$-$\sigma_8$ parameter space for the combined
shear-magnification arises because of the pre-factor $\Omega_m$ in
Eq. \ref{2ptstat}, while the shear scales as $\Omega_m^2$. In a
companion paper (\cite{hvwe2009}), we presents the first measurement
of cosmic magnification on galaxies. The source galaxies are
Lyman-break Galaxies (LBG), selected at high redshift using the
dropout technique as described in \cite{hh2009}. This technique
provides a robust way of separating background and foreground
galaxies. The signal was measured for a wide range of slopes
$\alpha$ which confirms the cosmological origin of the signal,
similar to \cite{scr2005}.

The main limitation for the use of cosmic magnification as a
cosmology probe is the explicit dependence on the galaxy biasing,
which in general cannot be assume to be $b=1$ or even constant with
scale. There is however a remarkable complementarity with cosmic
shear which can be exploited in order to measure the cosmological
parameters. The auto-correlation function of the foreground
population is indeed given by

\begin{equation}
\xi_N(\theta)=2b^2\int {\rm d}w~{p^2_f(w)\over f_K^2(w)}\int s{\rm
d}s~P_{3D}\left({s\over f_K(w)};w\right)~J_0(\theta s),
\end{equation}
and when combined with Eqs. \ref{2ptstat} it provides an independent
measurement of the usual shear two-points correlation function (Eq.
\ref{shear2pts}) proportional to $\propto \sigma_8^2 \Omega^2_m$,
which does not depend on galaxy biasing. The new estimator can be
written as:

\begin{equation}
\xi_\mu(\theta)={w_{12}^2(\theta)\over \xi_N(\theta)}.
\end{equation}
Such measurement uses photometry data only and does not rely on any
shape measurement. The approach is similar to \cite{w1998} who
showed that the combination of the shear with number counts provide
constraints on the galaxy biasing and cosmology. In that paper it
was shown that the equivalent of Eq. \ref{2ptstat} was the
cross-correlation between the shear of distant galaxies and the
number counts of the foreground galaxies:

\begin{eqnarray}
\langle M_{\rm ap}N_{\rm ap}\rangle&=&3\pi b~\Omega_m \int {s{\rm
d}s\over 2\pi}\int{\rm d}w {g(w)\over a(w)}{p_1(w)\over
f_K(w)}\times\cr && P_{3D}\left({s\over f_K(w)};w\right) I^2(\theta
s), \label{biasshear}
\end{eqnarray}
where $I(\theta s)$ is the Fourier transform of the aperture filter.
The first practical implementation combining shear and number counts
was very promising (\cite{hvwetal2002}).

The advantage of Eq. \ref{2ptstat} over Eq. \ref{biasshear} lies in
the fact that the cosmic magnification is completely independent of
residual systematics inherent to shear measurement. It shows that
the combined use of cosmic shear, magnification and number count
statistics could be optimized in order to measure the cosmological
parameters and simultaneously identify and reduce the systematics.
Among the systematics which affect the shear and not the
magnification one can mention the shear calibration and additive
bias, and the intrinsic alignment which couples galaxy orientation
over a large redshift range (\cite{hs2004}). High order shear
statistics are particularly sensitive to intrinsic alignment which
could eventually completely dominate the signal in some situations
(\cite{sem2008}).

Another nice feature of cosmic magnification is the possibility for
exploiting faint distant galaxies for weak lensing, galaxies for
which the shape cannot be measured, and therefore would not be used
otherwise. This is the case for instance for LBGs which can be
identified at very large redshift using the dropout technique in
optical bands. \cite{hvwe2009} have successfully measured the
magnification on redshift $z=3,4,5$ LBGs. This provides a very
interesting observational window for future deep large surveys such
as LSST and JDEM. With cosmmic shear measurement alone, a large
fraction of the detected objects in those surveys would not be used
for weak lensing studies. The combination of shear and magnification
enables the possibility to use all detected objects (including the
faint and/or high redshift galaxies and quasars) to probe dark
matter from weak lensing.

Like cosmic shear, the cosmic magnification can be measured at
various source redshifts, for various lens redshift, and for
different magnitude bins. A study which would combine shear and
magnification with a tomographic approach is left for a forthcoming
study.

\section{Conclusion}

This paper is a preliminary study of the cosmological applications
of the cosmic magnification effect. A general expression for the
covariance of the magnification estimator was derived, which is the
first step towards parameter measurement forecast. The CFHTLS Deep
survey was used to forecast the precision of magnification
measurements for future surveys and compared to the cosmic shear
measurements. Although cosmic magnification has a lower
signal-to-noise than cosmic shear, the former still contains useful
cosmological information which should be exploited in future
surveys, since to a large extend it will be obtained {\it for free}.
Note that the Eddington bias (\cite{edd1913}) has not been discussed
in this study, but it can be easily accounted for (\cite{hvwe2009}).

In many respects, shear and magnification appear as complementary
probes of the dark matter distribution. They are sensitive to
completely different observational systematics, therefore they can
be used as redundant measurements for a better control of the
residual systematics. Magnification is also not sensitive to
intrinsic alignment/shear correlations which is known to be a major
source of non lensing signal for shear tomography. Some techniques
have been proposed to remove intrinsic alignment using shear
information only (e.g. \cite{js2008}), cosmic magnification could
certainly help measuring it.

Although magnification is often presented as a difficult
measurement, it is certainly not more {\it difficult} than galaxy
shape measurement, moreover, future lensing surveys will have to
have high precision photometry in order to deliver precise
photometric redshifts. It is therefore natural to envision
forthcoming surveys combining both measurements, however it is not
clear at this stage that current plans for future surveys are yet
optimal for redshift measurements. The main source of systematics
for photometric redshift results from biases in the photometry. This
could come from 1) incorrect calibration 2) improper color coverage
or 3) small angular scale galactic and extragalactic dust
extinction. Point 1) could be addressed with space-based photometry
and/or partial spectroscopic followup. Point 2) could be easily
addressed with an increase of wavelength coverage and/or number of
filters. The importance of point 3) is unclear at the moment,
although recent results from the SLOAN suggest the effect is small
(\cite{mks2009}) and chromatic signature could certainly help to
disentangle weak lensing from dust extinction. At this stage,
further work is necessary in order to explore in more detail the
potential of cosmic magnification, in particular tomographic
magnification with varying lens and source redshifts look promising.
Only then, adjustments to future lensing missions could be
envisioned for a better photometric redshift measurements. The
explicit dependence of magnification on galaxy biasing is not a
major source of concern since it can in principle be constrained
from magnification tomography with varying source redshift and fixed
lens redshift, moreover the combination with galaxy angular
correlation functions and cosmic shear provides a direct measurement
of biasing.

One should note that the detection of magnification on LBGs
(\cite{hvwe2009}) is already a strong proof of concept, in that one
can now measure magnification on very high redshift galaxies, in a
regime where lensing is not sensitive to the exact value of $z_s$.
The background/foreground separation is also much easier to perform,
virtually free of any contamination, and error in the foreground
redshift distribution is probably not as problematic as if the
redshift bins were close. This would for instance make a strong case
for deep {\it u} or {\it g} coverage in order to identity redshift
$3$ and $4$ dropouts with better accuracy.

The measurement of weak lensing without relying on galaxy shapes
provides definitely an interesting complementary approach to cosmic
shear, although it is clear it will not {\it replace} it. The main
idea developed in this paper is that shear surveys could also
perform magnification measurements with minor upgrades in the
instrument design in order to make photometric redshift more
reliable, and the scientific gain is potentially substantial.
Current surveys will certainly significantly contribute in the
exploration of this new direction \footnote{CFHTLS, DES, PanSTARRS,
KIDS-VISTA}. It also enables wide field spectroscopic surveys such
as ADEPT to perform weak lensing measurements independently from
wide-field imaging surveys.

\section*{acknowledgements}
I would like to warmly thank Istvan Szapudi and Jonathan Benjamin
for discussions on angular auto and cross-correlation functions at
the beginning of this work. It is a pleasure to thank Hendrik
Hildebrandt, Martin White and Catherine Heymans for comments on the
manuscript. Support from NSERC, CIfAR and CFI is also acknowledged.

\end{document}